\documentclass[12pt, preprint]{aastex}
\usepackage{mathptmx, ifthen}
\usepackage{bm}
\usepackage{color} 
\shorttitle{Effects of Reduced Shear on Dark Energy Constraints}
\shortauthors{Shapiro}

\def\eg{{\it e.g.} }  
\def\ie{{\it i.e.} }
\def\etal{{\it et al.} }

\def\LCDM{$\Lambda$CDM~}
\newcommand{\Mpc}{\mbox{ Mpc}}
\newcommand{\Mpch}{\mbox{ Mpc}/h}
\newcommand{\hMpc}{h \mbox{ Mpc}^{-1}}
\newcommand{\fsky}{f^{\rm sky}}
\newcommand{\ngal}{n^{\rm gal}}
\newcommand{\chieff}{\chi^{\rm eff}}

\newcommand{\ishear}{\gamma_{\rm rms}}
\newcommand{\zmed}{z^{\rm med}}

\def\vl{\bm l} 
\def\vt{\bm\theta}
\def\vx{\bm x}

\newcommand{\vk}[1]{\bm k_{#1}}
\newcommand{\dd}[1]{d #1\,}
\newcommand{\Clmat}{{\mathbf C}_l}
\newcommand{\trace}{\mbox{Tr}}
\newcommand{\dzeta}{\delta_\zeta}
\newcommand{\ten}[1]{\times 10^{#1}}
\newcommand{\shearrat}[3]{\Gamma_{#1}(#3;#2)}
\newcommand{\Fw}{F^{\rm w}}

\newcommand{\mean}[1]{\left\langle{#1}\right\rangle}
\newcommand{\partder}[2]{\frac{\partial #1}{\partial #2}}
\def\be{\begin{equation}}
\def\ee{\end{equation}}
\def\bea{\begin{eqnarray}}
\def\eea{\end{eqnarray}}

\newcommand{\refeq}[1]{(\ref{eq:#1})}
\newcommand{\reftab}[1]{Table \ref{tab:#1}}
\newcommand{\reffig}[1]{Figure \ref{fig:#1}}

\begin{document}

\title{Biased Dark Energy Constraints from Neglecting Reduced Shear in Weak Lensing Surveys}
\author{Charles Shapiro}
\date{Draft as of \today}

\affil{Department of Physics, Enrico Fermi Institute \\ The University of Chicago, Chicago, IL 60637}
\affil{Institute of Cosmology and Gravitation \\ University of Portsmouth \\ Portsmouth, UK, PO1 2EG}
\email{charles.shapiro@port.ac.uk}

\begin{abstract}
The weak gravitational lensing of distant galaxies by large-scale structure is expected to become a powerful probe of dark energy.  By measuring the ellipticities of large numbers of background galaxies, the subtle gravitational distortion called ``cosmic shear'' can be measured and used to constrain dark energy parameters.  The observed galaxy ellipticities, however, are induced not by shear but by {\em reduced shear}, which also accounts for slight magnifications of the images.  This distinction is negligible for present weak lensing surveys, but it will become more important as we improve our ability to measure and understand small-angle cosmic shear modes.  I calculate the discrepancy between shear and reduced shear in the context of power spectra and cross spectra, finding the difference could be as high as 10\% on the smallest accessible angular scales.  I estimate how this difference will bias dark energy parameters obtained from two weak lensing methods: weak lensing tomography and the shear ratio method known as offset-linear scaling.  For weak lensing tomography, ignoring the effects of reduced shear will cause future surveys considered by the Dark Energy Task Force to measure dark energy parameters that are biased by amounts comparable to their error bars.  I advocate that reduced shear be properly accounted for in such surveys, and I provide a semi-analytic formula for doing so.  Since reduced shear cross spectra do not follow an offset-linear scaling relation, the shear ratio method is similarly biased but with smaller significance.
\end{abstract}





\section{Introduction}
\label{cha:intro}

Weak gravitational lensing of galaxies has rapidly become an essential cosmological tool.  Cosmic shear -- the large-scale pattern of gravitational distortion embedded in the observed shapes of galaxies -- is a particularly valuable effect that contains information about the growth of mass fluctuations as well as the Universe's expansion history and spatial curvature.  Cosmic shear was first detected by multiple independent groups at the turn of the century \citep{bacon_refregier_etal_2000, kaiser_wilson_etal_2000,van-waerbeke_mellier_etal_2000, wittman_tyson_etal_2000}, and it could become one of the premier methods for studying dark energy \citep{albrecht_bernstein_etal_2006}.  Subsequent surveys, such as the 100 Square Degree Weak Lensing Survey, have shown that cosmic shear measurements provide cosmological parameter constraints that are generally consistent with more mature observations such as CMB anisotropy, galaxy correlations, and galaxy cluster counting \citep{benjamin_heymans_etal_2007}.  This survey was a compilation of four separate surveys which, together, provided 3.5 million galaxies with a median redshift of about $\zmed\sim 0.8$.  Ongoing surveys and those on the horizon are expected to provide wider, deeper multi-color images needed to extract information about dark energy.  For example, NASA's Joint Dark Energy Mission is a planned space-based telescope that would survey 4000 deg$^2$, measuring shapes and photometric redshifts for over a {\em billion} galaxies with a median redshift of $\zmed\sim 1.5$.

As daunting as the observational systematics are for ambitious shear surveys, there are also important theoretical hurdles that must be overcome so that our ability to predict the shear signal will keep pace with our ability to measure it.  Various theoretical systematics (uncertainties or complications in the predictions) have been identified and in some cases addressed.  Determining how baryons and non-linear gravitational clustering affect the matter power spectrum on $\sim 1 \Mpch$ scales is a primary concern \citep{huterer_takada_2005, rudd_zentner_etal_2008}.  Significant theoretical systematics also include the effects of non-gaussianity \citep{cooray_hu_2001} and the spurious signals arising from the intrinsic alignment and clustering of source galaxies \citep{schneider_van-waerbeke_etal_2002, hamana_colombi_etal_2002, heymans_heavens_2003, hirata_seljak_2004}.  Computational approximations are also being investigated more closely as we start to test regimes where simple first order shear calculations may no longer hold \citep{vale_white_2003, cooray_hu_2002, shapiro_cooray_2006, dodelson_kolb_etal_2005, dodelson_zhang_2005}.

Ignoring the difference between shear and reduced shear is one example of a  computational approximation that will not be accurate enough for upcoming cosmic shear surveys \citep{dodelson_shapiro_etal_2006}.  When a galaxy is weakly lensed, the change in ellipticity of its image is proportional to the {\em reduced shear}\,:
\begin{equation} \label{eq:defineRS}
g^a \equiv \frac{\gamma^a}{1-\kappa}
\end{equation}
where $\gamma$ is lensing shear and $\kappa$ is lensing convergence, a measure of the matter fluctuations projected along the line of sight.  The superscripts denote a ``+'' or ``$\times$'' shear polarization transverse to the line of sight.  When the distortion fields $\kappa$ and $\gamma$ are much less than unity, we can expand \refeq{defineRS} to first order in the fields, making the approximation that $g^a\approx\gamma^a$, \ie that galaxy ellipticities directly measure the shear field.  This approximation is widely used by theorists since it holds for the vast majority of galaxy images in a cosmic shear survey and since it makes calculating the cosmic shear power spectrum and bispectrum much more manageable.  Dodelson, Shapiro, and White (\citeyear{dodelson_shapiro_etal_2006}) have found that this seemingly innocuous substitution can nevertheless change the cosmic shear spectrum by as much as 10\% on arcminute angular scales, where lensing is enhanced by the non-linear matter power spectrum.  Thus, ignoring reduced shear in this regime could bias a survey's analysis, causing it to rule out a correct set of cosmological parameters.

This paper assesses the impact of the reduced shear approximation on dark energy parameter determination from cosmic shear.  Dodelson \etal considered source galaxies at a single redshift, and hence only assessed the bias that reduced shear would have on \LCDM models.  Here I generalize to multi-redshift weak lensing analyses that constrain dark energy models by using information about distances and/or the growth of massive structures.  The two methods I consider are weak lensing tomography and a shear ratio method.  Tomography is more well-known: using the CDM paradigm combined with some dark energy model, one calculates a prediction for cosmic shear observables for sources in multiple redshift bins, and then the predictions are compared with observation; model parameters can subsequently be fit to the data \citep{hu_2002, albrecht_bernstein_etal_2006}.  Shear ratio methods forgo theoretical predictions of the shear signal and attempt to extract model-independent distance measurements by taking ratios of the shear signal at different redshifts.  I will focus on the ``offset-linear scaling'' shear ratio method of \citet{zhang_hui_etal_2005}, which is an alternative to the ``cross-correlation cosmography'' method of \citet{jain_taylor_2003}.

The structure of this paper is as follows.  In section 2, I will review some relevant weak lensing formalism and discuss how the next order correction to \refeq{defineRS} alters cosmic shear power spectra.  In section 3, I estimate  the bias in dark energy parameters that various surveys would experience by using the reduced shear approximation with weak lensing tomography.  I review the offset-linear scaling method in section 4 and similarly calculate how it is biased by reduced shear.

\section{The Reduced Shear Power Spectrum}

\subsection{Review of the leading order result}
We begin with a review of relevant weak lensing equations \citep[for a full introduction see \eg][]{blandford_saust_etal_1991, kaiser_1992, dodelson_2003, bartelmann_schneider_2001}.  The lensing convergence at a particular sky position, $\kappa(\vt)$, can be expressed as the matter density contrast, $\delta(\vx )$, projected over comoving distance, $\chi$, along the line of sight:
\be
\kappa_i(\vt)=\int_0^\infty d\chi\,\delta(\vt\chi,\chi)\,W_i(\chi) .
\ee
The subscript denotes a particular redshift bin $i$ from which source galaxies have been selected, and $W_i(\chi)$ is the lensing kernel for sources in the $i$th redshift bin, defined below.  Future cosmic shear surveys will measure galaxy redshifts photometrically, estimating each redshift by observing the galaxy's flux in a few broad wavelength bands.  The alternative is to obtain a redshift from the spectrum of each galaxy; although spectroscopic redshifts are much more precise, they will become prohibitively expensive and time-consuming for surveys of many millions of galaxies.
If the source galaxies are binned according to their photometric redshifts, 
then
\be
W_i(\chi) = 
\frac{W_0}{\ngal_i}\frac{d_A(\chi)}{a(\chi)}\int_{\chi}^{\infty} \dd{\chi_s} p_i(z)\frac{dz}{d\chi_s}\frac{d_A(\chi_s-\chi)}{d_A(\chi_s)}
\ee
with $W_0=\frac{3}{2}\Omega_m H_0^2$.  Here, $p_i(z)$ is the true (spectroscopic) distribution of galaxies in the $i$th photometric redshift bin, and $\ngal_i$ is the total 2D number density of galaxies in that bin.  The scale factor relative to today is $a\equiv(1+z)^{-1}$, and $d_A$ is a function that accounts for spatial curvature in open or closed Universes:
\begin{equation}
d_A(\chi) \equiv \left\{
\begin{array}{cccc}
\chi &{\rm for}& K=0 & {\rm (flat)} \\
|K|^{-1/2}  \sin(|K|^{1/2}\chi) &{\rm for}& K>0 &{\rm (closed)}\\
|K|^{-1/2}  \sinh(|K|^{1/2}\chi) &{\rm for}& K<0 & {\rm (open)}
\end{array}
\right.
.
\end{equation}

To leading order, the two shear fields $\gamma^a(\vt)$ are easily related to the convergence in Fourier space:
\be
\tilde\gamma^a_i(\vl) = T^a(\vl)\tilde\kappa_i(\vl)
\: ,
\ee
where $\vl$ is the Fourier conjugate of $\vt$, and I am working in the small angle limit, rather than decomposing the fields into spherical harmonics.  The trigonometric weighting functions are
\be
T^1(\vl)\equiv \cos(2\phi_l) \hspace{1cm}
T^2(\vl)\equiv \sin(2\phi_l)
\ee
where $\phi_l$ is the angle between $\vl$ and some fixed $x$-axis.  It is useful to consider the following linear combinations of the shear modes,
\bea
\tilde E_i(\vl) &=& T^a(\vl) \tilde\gamma^a_i(\vl) \label{eq:defE}\\
\tilde B_i(\vl) &=& \epsilon^{ab} T^a(\vl) \tilde\gamma^b_i(\vl)
\: , \eea
where $\epsilon^{ab}$ is the anti-symmetric matrix, $\epsilon^{12}=-\epsilon^{21}=1$.  To leading order, the E-mode and the convergence have the same fourier coefficients, $\tilde E_i(\vl)=\tilde\kappa_i(\vl)$, while the B-mode vanishes.  These relations do not account for small, second-order corrections to the weak lensing fields such as lens-lens coupling or source clustering \citep{cooray_hu_2002, schneider_van-waerbeke_etal_2002, hamana_colombi_etal_2002}.  The E-mode angular power spectra and cross spectra, $C_{l;ij}$, are defined as
\be
\mean{\tilde E_i(\vl) \tilde E_j(\vl')}
\equiv (2\pi)^2 \delta^2\!\left(\vl+\vl'\right) C_{l;ij}
\ee
where $\delta^2$ is a 2-dimensional Dirac delta function and the angle brackets denote an ensemble average.  The leading order calculation of the $C_{l;ij}$ is well known:
\be \label{eq:shearCl}
C_{l;ij} = \int_0^\infty \frac{\dd{\chi}}{d_A(\chi)^2} W_i(\chi)W_j(\chi) P_\delta\left(k;\chi\right)
\ee
where $P_\delta\left(k;\chi\right)$ is the 3D matter power spectrum for $k=\ell/d_A(\chi)$ at a distance $\chi$, accounting for the growth of structure.  The above expression uses the Limber approximation, which assumes that the only matter density modes $\tilde\delta(\vk{})$ contributing to the lensing signal are those modes with $\vk{}$ transverse to the line of sight.  Equation \refeq{shearCl} is commonly used to predict the cosmic shear signal in order to constrain cosmological parameters.  In particular, dark energy can be constrained via its effects on structure growth and on the geometrical functions, $W_i(\chi)/d_A(\chi)$.

\subsection{Perturbative Correction}

Since galaxy shape measurements are sensitive not to shear but to reduced shear, we must consider the difference between the shear spectra in \refeq{shearCl} and the expected reduced shear spectra.  We can do this perturbatively by expanding \refeq{defineRS} about $\kappa=0$ and keeping one more order:
\begin{eqnarray}
\label{eq:expandRS}
g^a_i(\vt) &=& \gamma^a_i(\vt) + (\gamma^a\kappa)_i(\vt) + \ldots \\
& & \mathcal{O}(\kappa^1) \hspace{.6cm}\mathcal{O}(\kappa^2) \hspace{.6cm}\mathcal{O}(\kappa^3)\nonumber
\end{eqnarray}
If one then recomputes the E-mode spectra, substituting $\tilde{g}$ for $\tilde{\gamma}$ in \refeq{defE}, one recovers the leading order spectrum plus a correction term:
\be
\delta \mean{\tilde E_i(\vl) \tilde E_j(\vl')}
= T^a(\vl) T^b(\vl')
\mean{\widetilde{(\gamma^a\kappa)}_i(\vl)\, \tilde{\gamma}_j^b(\vl')}
+ \left( \vl\leftrightarrow \vl' \right)
\: \ee
Under the Limber approximation, the power- and cross- spectra corrections from reduced shear simplify to an easily calculable form
\footnote{The corresponding equation in \citet{dodelson_shapiro_etal_2006} is incorrect as printed, containing spurious factors of $\vl$ and $\vl'$.  It should match \refeq{RScorrect} here with $i=j$.  Here I use different conventions for the $T^a(\vl)$.}
: 
\be \label{eq:RScorrect}
\delta C_{l;ij}=2 \int \frac{d^2l'}{(2\pi)^2}\,\cos(2\phi_{l'}-2\phi_l)
 \,B^\kappa_{ij}(\vl,\vl',-\vl-\vl';\chi)
\ee
where I have defined the 2-redshift convergence bispectrum,
\be
B^\kappa_{ij}(\vl_1,\vl_2,\vl_3;\chi)\equiv
\frac{1}{2} \int_0^\infty \frac{d\chi}{d_A(\chi)^4}W_i(\chi)W_j(\chi)[W_i(\chi)+W_j(\chi)]
\, B_\delta\left(\vk{1},\vk{2},\vk{3};\chi\right)
\ee
with no sum over $i$ or $j$ and e.g. $\vk{1}=\vl_1/d_A(\chi)$.  Thus, the reduced shear correction is proportional to a 2D convergence bispectrum, which can be written
as a projection of the 3D matter bispectrum, just as the leading order shear power spectrum is a projection of the matter power spectrum.  Due to the form of the matter bispectrum \citep{scoccimarro_couchman_2001} the integrand in \refeq{RScorrect} depends only on $\vl\cdot\vl'$, and therefore $\delta C_{l;ij}$ is independent of the direction of $\vl$.  Under the Limber approximation, I find no corrections to the $\tilde{E}\tilde{B}$ or $\tilde{B}\tilde{B}$ spectra to 3rd order in the weak lensing fields.

In 2nd order perturbation theory the matter bispectrum, $B_\delta$, is proportional to $P_\delta^2$; using this result as a starting point, an accurate fitting formula for $B_\delta$ was obtained from N-body simulations by \citet{scoccimarro_couchman_2001}.  I use their formula to compute the reduced shear correction, $\delta C_{l;ij}$.  I compute the linear matter power spectrum using the fitting formula of \cite{eisenstein_hu_1999}, and the non-linear part is computed from the ``Halofit'' code of \cite{smith_peacock_etal_2003} which has also been calibrated by numerical simulations.  This is the same procedure used by \cite{dodelson_shapiro_etal_2006}, who demonstrated that the semi-analytic formula for the reduced shear correction \refeq{RScorrect} adequately agrees with the difference found by ray-tracing N-body simulations (for all source galaxies at $z=1$).  Encouraged by this agreement, and because other studies have shown that corrections of order $\kappa^4$ are negligible \citep{cooray_hu_2002, shapiro_cooray_2006, krause_2008}, I do not refine the correction any further by including higher order terms in \refeq{expandRS}.

\begin{figure}
\begin{center}
\includegraphics[width=5.75in]{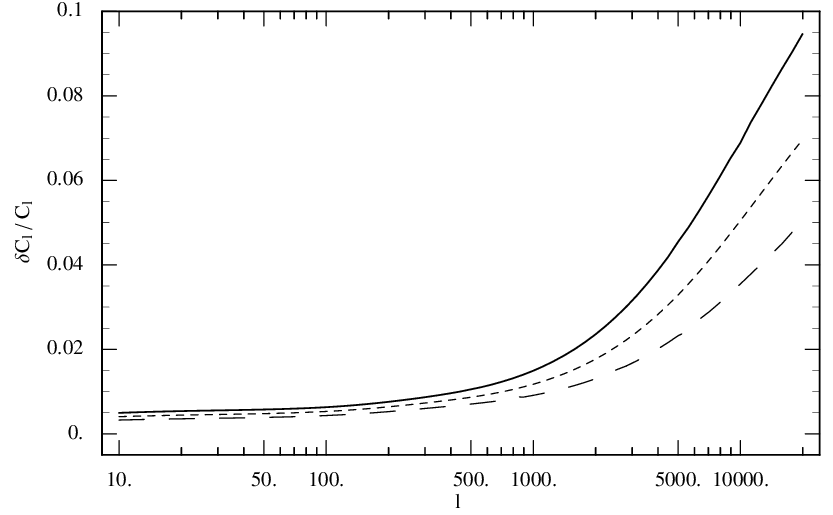}
\end{center}
\caption[Reduced shear correction, dependence on $\sigma_8$] {Dependence of the reduced shear correction on $\sigma_8$.  The size of the correction relative to the leading order cosmic shear power spectrum is shown as a function of multipole.  From smallest to largest, the curves assume $\sigma_8$ = 0.7, 0.8, and 0.9.  All curves are for source galaxies in a single redshift bin with $0.8<z<1.0$.
\label{fig:sigma8plot}  }
\end{figure}

\begin{figure}
\begin{center}
\includegraphics[width=5.3in]{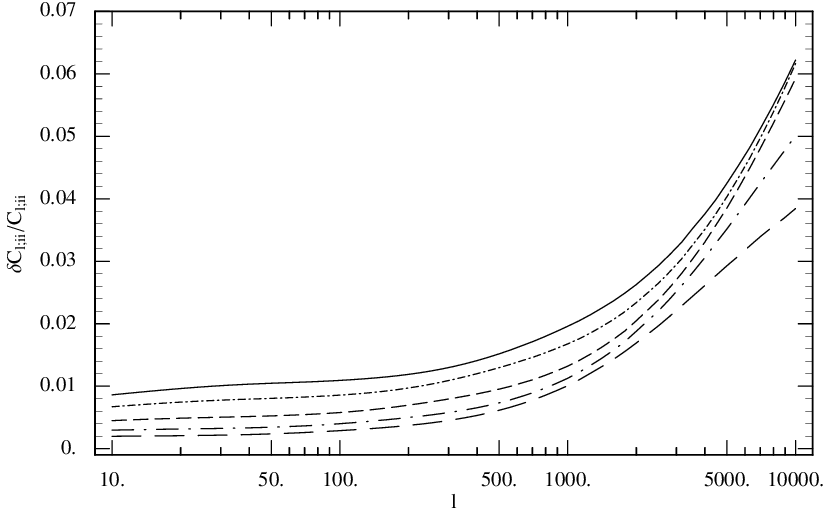}
\includegraphics[width=5.3in]{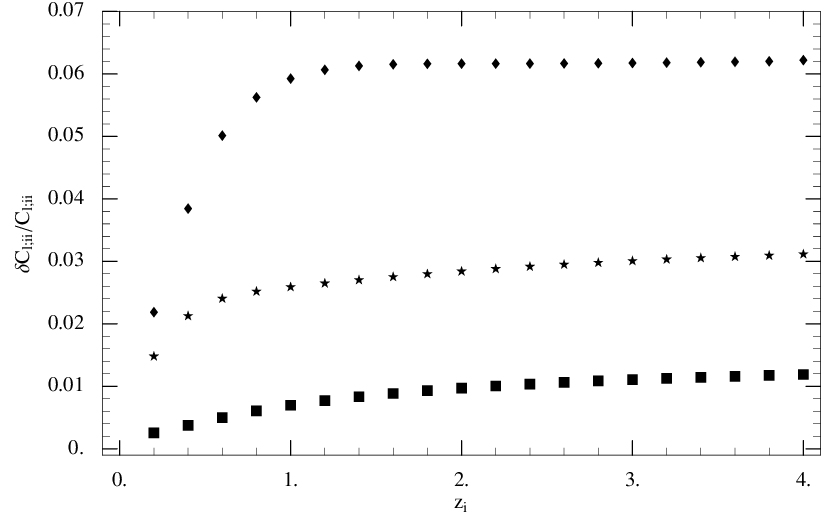}
\end{center}
\caption[Reduced shear correction, dependence on redshift and multipole] {Dependence of the reduced shear correction on redshift and multipole.  Top: same as figure \ref{fig:sigma8plot} but for different redshift bins and $\sigma_8$ fixed at 0.85.  For source galaxies in a single redshift bin of width $\Delta z=0.2$, the curves are for bins with maximum redshifts $z_i$ = 0.4, 0.6, 1.0, 2.0, 4.0.  The correction increases with increasing redshift.  Bottom: same correction as a function of $z_i$.  From smallest to largest, the curves fix $l=$ 200, 2820, 10,000.
\label{fig:RSvZ}  }
\end{figure}

Figures \ref{fig:sigma8plot} and \ref{fig:RSvZ} show the size of the reduced shear correction relative to the leading order cosmic shear power spectrum.  Unless otherwise noted, all calculations use the reference \LCDM cosmological model summarized in \reftab{cosmopars}.
The correction is most significant toward the non-linear part of the shear spectrum, reaching several percent on small scales corresponding to lensing by individual dark matter halos.  Essentially, the correction reflects the ratio of the matter bispectrum to the matter power spectrum, which is enhanced by non-linearity; and indeed when the computation is repeated using only linear matter fluctuations, the correction is negligible on all scales.  Thus it is not surprising that the size of the correction is sensitive to a parameter like the matter power spectrum normalization, $\sigma_8$, which strongly determines the matter power spectrum on non-linear scales.  As shown in figure \ref{fig:sigma8plot}, the correction is enhanced by a high value of $\sigma_8$, and it would be similarly enhanced by raising the primordial spectral tilt $n_s$.  We can see that the correction could reach the 10\% level for $l=2\ten{4}$, but beyond this scale, shape noise will dominate even the deepest surveys of the near future.

The reduced shear correction also increases with redshift, as shown in Figure \ref{fig:RSvZ}.  This is expected on all scales since the lensing kernel $W_i(\chi)$ increases with redshift, and the convergence bispectrum has an extra factor of $W_i(\chi)$ relative to the power spectrum.  Note that while the  correction increases steadily on large scales, it quickly approaches a maximum on small scales.  This reflects the fact that the small-scale bispectrum to power spectrum ratio drops off dramatically in the more linear, high-redshift Universe.  In other words, light rays encounter the most non-linear structure at low redshifts.  Again, when these computations are repeated using only the linear matter power spectrum, the correction at $l=$ 10,000 rises steadily with redshift instead of flattening out. 

\section{Implications for Dark Energy Constraints from Tomography}
\label{tomography}

\subsection{Methodology}

Having calculated the difference that reduced shear makes in the expected cosmic shear power- and cross- spectra, we would like to know to what extent ignoring this difference will bias measurements of cosmological parameters.  This subsection focuses on the technique of weak lensing tomography, in which measurements of cosmic shear spectra are simply compared to the predictions of some model and then used to fit the model's parameters.  To estimate the bias in a set of parameters, I first compute a Fisher matrix \citep[see e.g.][]{dodelson_2003} which approximates the expected inverse covariance matrix for those parameters:
\begin{equation} \label{eq:Fisher}
F_{\alpha\beta}=
\frac{1}{2}\sum_l(2l+1)
\trace \left(
\Clmat^{-1} \cdot \partder{\Clmat}{p_\alpha} \cdot \Clmat^{-1} \cdot \partder{\Clmat}{p_\beta}
\right )
\end{equation}
where $p_\alpha$ is the value of the $\alpha$th parameter, and the $\Clmat$ are the symmetric covariance matrices of the shear signal.  I am assuming that the various modes $l$ are uncorrelated -- this is an approximation discussed in more detail at the end of this section.  The elements of $\Clmat$ are the expected shear signal plus shape noise, $C_{l;ij}+N_{ij}$.  The shape noise matrix is diagonal and accounts for the fact that the shear in redshift bin $i$ is measured from a finite number density of background galaxies $\ngal_i$ with intrinsic shape variance $\ishear^2$:
\begin{equation}
N_{ii}=\frac{\gamma_{\rm rms}^2}{\ngal_i}
\:. \end{equation}
If prior constraints on the parameters are included, the total Fisher matrix is simply the sum of the prior matrix and the survey matrix:
\begin{equation} \label{eq:FisherTot}
F^{\rm tot}_{\alpha\beta} = \fsky F_{\alpha\beta} + F^{\rm prior}_{\alpha\beta}
\end{equation}
where $\fsky$ is the sky coverage of the survey, expressed as a fraction of the total sky.  Diagonal elements of the inverse matrix, $(F^{\rm tot})^{-1}$, are estimates of the forecasted variances of the parameters, marginalized over all other parameters.  These estimates are accurate when the signal and noise distributions are approximately Gaussian.

\begin{figure}
\begin{center}
\includegraphics[width=4.0in]{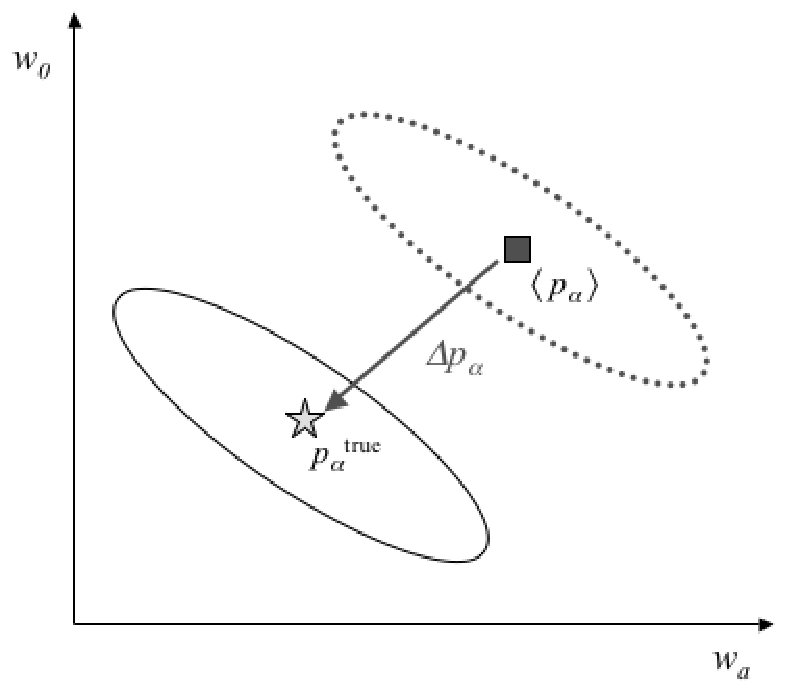}
\end{center}
\caption[Illustration of parameter bias $\Delta p_\alpha$] {Illustration of parameter bias $\Delta p_i$ in the $w_0-w_a$ plane.  The star represents the true parameter values, and the solid ellipse represents the 1$\sigma$ error contour forecasted for an unbiased survey via a Fisher matrix analysis.  The square and dotted ellipse represent the best-fit parameters and 1$\sigma$ error contour expected from a survey that ignores reduced shear.  In this example, the best-fit dark energy parameters expected from the survey are incorrect by several ``sigma''. 
\label{fig:FoBdiagram}  }
\end{figure}

If our predictions of the shear signal are incorrect, then we expect there to be a bias in $\mean{p_\alpha}$, the expected best-fit parameter set from a survey using those predictions.  Now from \refeq{Fisher} and \refeq{FisherTot} it is straightforward to show that if the true cosmology is given by the parameters $p^{\rm true}_\alpha$, then to first order, the parameter differences are given by
\begin{equation}
\Delta p_\alpha \equiv p_\alpha^{\rm true}-\mean{p_\alpha}=
\frac{\fsky}{2}(F^{\rm tot})_{\alpha\beta}^{-1}\sum_l(2l+1)
\trace \left(
\Clmat^{-1} \cdot \partder{\Clmat}{p_\beta} \cdot \Clmat^{-1} \cdot \delta\Clmat
\right )
\end{equation}
where the $\delta \Clmat$ are matrices whose elements are the ``prediction errors'', in this case, the reduced shear corrections $\delta C_{l;ij}$.  The concern is then whether this bias $\Delta p_\alpha$ is large compared to the survey's confidence intervals about $\mean{p_\alpha}$, as illustrated in \reffig{FoBdiagram}.

If the parameter biases $\Delta p_\alpha$ are substantial relative to the parameter confidence intervals obtained from the Fisher matrix, then we should be concerned about ignoring the difference between shear and reduced shear.  However, 1-dimensional biases and error bars do not provide the full picture: it is possible to slightly bias multiple parameters so as to cause their {\em combination} to be ruled out in the full multi-dimensional parameter space.  In particular, I am interested in the dark energy equation of state parameters $w_0$ and $w_a$, defined by
\begin{equation}
w(z) = w_0 + w_a(1-a) = w_0 + w_a\; \frac{z}{1+z}
. \end{equation}
To check whether these two parameters will be biased in combination, I have calculated a ``figure of bias'' (FoB) where appropriate \citep[a similar metric is used in][]{dodelson_shapiro_etal_2006}.

To obtain the FoB, I first invert the Fisher matrix $F^{\rm tot}$ to obtain a covariance matrix for the cosmological parameters.  I then extract the $w_{0}-w_{a}$ block of that covariance matrix and call this submatrix $(\Fw)^{-1}$.  The FoB is defined as
\begin{equation}
\mbox{FoB} \equiv \left(\sum_{\alpha\beta}\, \Delta p_\alpha \Fw_{\alpha\beta}\, \Delta p_\beta\right)^{1/2}
\end{equation}
where the sum only runs over the parameters $w_0$ and $w_a$.
Note that for a single parameter, the FoB is simply $\Delta p/\sigma(p)$.  More generally, the FoB indicates the level at which we should expect a survey to rule out the correct dark energy parameter combination due to the reduced shear bias.  With no bias, we expect (in a statistical sense) that the $w_{0}-w_{a}$ confidence region will be centered on the true dark energy values like the solid ellipse in \reffig{FoBdiagram}.  With a bias, we expect that confidence region to be displaced by the vector $\Delta p_\alpha$, thereby placing the true parameters on some error contour.  Assuming that the error contours are ellipses defined by the Fisher matrix $\Fw$, the FoB squared is simply the $\Delta\chi^2$ of the true dark energy point relative to the center of the ellipses \citep[][see discussion of confidence intervals]{press_teukolsky_etal_1992}.  Some common confidence levels are given here:
\[
\begin{array}{ccccccc}
\hline
\mbox{FoB} 			& 1.52 & 2.15 & 2.48 & 3.03 & 3.44 & 4.29 \\
\mbox{(FoB)$^2$} 			& 2.30 & 4.61 & 6.17 & 9.21 & 11.8 & 18.4 \\
\mbox{Error contour}& 68.3\% & 90\% & 95.4\% & 99\% & 99.73\% & 99.99\% \\
\hline
\end{array}
\]
Thus if FoB = 1.52, it means that we expect the true $w_{0}$ and $w_{a}$ values to lie on the 68.3\% error contour of a survey's biased confidence region; FoB = 3.03 means that the true dark energy parameters would be ``ruled out'' at the 99\% level.

In order to estimate the impact of the reduced shear approximation on parameter fitting, I compute the Fisher matrix for three types of weak lensing surveys described in the Report of the Dark Energy Task Force \citep{albrecht_bernstein_etal_2006}.  The types of survey instruments I consider are a Stage-III 4-m class telescope using photometric redshifts, a Stage-IV large survey telescope (LST), and a Stage-IV Joint Dark Energy Mission (JDEM) satellite telescope.
All surveys are assumed to collect a population of source galaxies with a redshift distribution given by
\begin{equation}
\frac{d\ngal}{dz} \propto z^2 \exp[-(z/z_0)^{1.5}]
\:. \end{equation}
The median redshift of this distribution is $z_{\rm med}=z_0\sqrt{2}$, and the distribution is normalized so that the total angular number density of galaxies is
\begin{equation}
\int_0^\infty dz\,\frac{d\ngal}{dz}=\sum_i\ngal_i \equiv\ngal
\:. \end{equation}
I assume that photometric redshift measurements for all surveys are unbiased and have a Gaussian distribution about the true, spectroscopic galaxy redshifts with a scatter of $\sigma(z^{\rm phot})=0.02(1+z)$ \citep[for photo-z methodology see e.g.][]{ma_hu_etal_2006}.  To avoid degrading parameter constraints, it is actually more important to have a small uncertainty in the scatter rather than a small scatter \citep{ma_hu_etal_2006}.  The galaxies are divided into five photometric redshift bins, defined so as to have precisely equal shape noise when $\sigma(z^{\rm phot})=0$.  Several authors have found that subdividing the redshifts further, which increases the shape noise in each bin, does not significantly improve parameter constraints \citep{hu_1999,ma_hu_etal_2006,takada_jain_2004}.  These and other survey characteristics are summarized in \reftab{surveychars}.
\begin{table}[tdp]
\caption[Characteristics for three cosmic shear surveys]{Characteristics for three cosmic shear surveys suggested by the Dark Energy Task Force \citep{albrecht_bernstein_etal_2006}.}
\begin{center}
\begin{tabular}{@{} ccccl @{}}
\hline 
    Parameter 	& 4-m & LST & JDEM & Description \\
\hline
    $\fsky$		& 0.1 & 0.5 & 0.1	& Sky coverage of survey \\
    $\ngal$		& 15 & 40 & 100 	& Effective number of galaxies per arcmin$^2$\\
    $\zmed$& 1 & 1 & 1.5 		& Median redshift of galaxies \\
    $\ishear$& 0.25 & 0.25 & 0.3& Intrinsic galaxy shape variability \\
    $\sigma(f^{\rm cal})$& 0.01 & 0.01 & 0.001 & Multiplicative shear calibration error \\
\hline\hline
\\
\multicolumn{2}{l}{Photo-z bin edges}&\multicolumn{2}{l}{4-m, LST:}& 0, 0.622, 0.875, 1.13, 1.47, 4 \\
 & & \multicolumn{2}{l}{JDEM:} &  0, 0.930, 1.31, 1.69, 2.19, 4 \\
 \\
\hline
\end{tabular}
\end{center}
\label{tab:surveychars}
\end{table}

\begin{table}[htdp]
\caption[Cosmological parameter set and fiducial values]{Cosmological parameter set and fiducial values used to predict constraints and biases from weak lensing tomography.  The last group of parameters are not independent of the others.  The Planck Prior column contains 1$\sigma$ error bars derived from the Planck Fisher matrix.  The Planck constraints on curvature and dark energy parameters contain a degeneracy that precludes finite error bars.}
\begin{center}
\begin{tabular}{@{} cccl @{}}
\hline 
    Parameter 	& Fiducial	& Planck	& Description \\
		& Value		& Prior		&		\\
\hline
    $w_0$	& -1		&  		& Dark energy equation of state today \\
    $w_a$	& 0		&		& High-$z$ change in dark energy equation of state \\
    $\Omega_{DE}$& 0.72		&		& Dark energy density today \\
    $\Omega_k$	& 0			&		& Spatial curvature \\
    $\omega_m$	& 0.137		& 0.0012	& Total matter density today, $\Omega_m h^2$ \\
    $\omega_b$	& 0.0225	& 0.00017	& Baryon density today, $\Omega_b h^2$ \\
    $n_s$	& 0.96		& 0.0062		& Primordial scalar spectral index \\
    $\ln A_s$	& -0.249	& 0.018		& Log amplitude of primordial scalar spectrum \\
		&		&		& at $k=0.05\Mpc^{-1}$ \\
\hline
    $h$		& 0.7		&		& Hubble constant, $H_0\equiv 100\,h \mbox{ km/s/Mpc}$ \\
    $\sigma_8$	& 0.85		&		& Amplitude of linear matter power spectrum \\
		&		&		& on $8 \hMpc$ scales today \\
	$\dzeta^2$& $2.3\ten{-9}$&   		& Scalar amplitude, $\dzeta^2 = A_s\times 2.95\ten{-9}$ \\
\hline
\end{tabular}
\end{center}
\label{tab:cosmopars}
\end{table}

The cosmological model used for forecasting is summarized in \reftab{cosmopars}; it is a standard set of parameters consistent with WMAP 5-year data \citep{dunkley_komatsu_etal_2008}.  Curvature and the dark energy equation of state are allowed to vary, and fiducially the model is flat with a cosmological constant ($w_0=-1, w_a=0$).  I have intentionally chosen a somewhat high-end value of $\sigma_8$ that is pessimistic, \ie it increases the importance of the reduced shear correction, as shown in \reffig{sigma8plot}; this choice will make the final results more conservative.  I impose priors expected from CMB temperature and polarization measurements from the Planck satellite; this is accomplished by adding Planck's Fisher matrix for the parameters in \reftab{cosmopars} to the Fisher matrix of each weak lensing survey.  The Planck Fisher matrix was computed by the Dark Energy Task Force and was made available with the interactive program, ``DETFast'' \citep{dick_detfast}.  In addition, I marginalize over the nuisance parameters $f^{\rm cal}_{i}$ to account for multiplicative shear calibration errors in each redshift bin: $C_{l;ij}\propto f^{\rm cal}_{i}f^{\rm cal}_{j}$.  These parameters are equal to unity, with errors given in \reftab{surveychars}.

The shear modes included in my analysis are $10\le l\le 3000$; a high-$l$ cutoff is necessary since baryons and non-linear gravitational clustering make the shear fields on smaller scales non-Gaussian and difficult to predict as of yet \citep{hu_white_2001, rudd_zentner_etal_2008}.  Since the $C_{l;ij}$ are rather smooth functions of $l$, in practice I compute sums by sampling logarithmically in $l$, and I have checked that the sampling is fine enough for the results to have converged.  Although I have treated the shear modes as independent, non-linear clustering does in fact correlate the matter density modes $\delta(\vk{})$, thereby correlating the shear \citep{hu_white_2001, rimes_hamilton_2005}.  These correlations can degrade parameter constraints (depending on the parameter), and so can non-Gaussianity; therefore the parameter errors obtained from the Fisher matrix will be lower limits.  I numerically compute two-sided derivatives of the observables with step sizes of 0.1 for $w_0$ and $w_a$, .025 for $n_s$ and $\ln A_s$, and 5\% for the remaining parameters; again, I check that reducing the step size does not significantly affect the results.

\newpage
\subsection{Results and Discussion}

Forecasted cosmological parameter constraints and biases for each of the three surveys considered are shown in Tables \ref{tab:DEStable}, \ref{tab:LSSTtable}, and \ref{tab:SNAPtable}.
In the case of the 4-m class telescope, we can see that the dark energy biases are comparable to their 1$\sigma$ error bars.  The equation of state parameters are biased by more than 0.5$\sigma$ in all cases, and the dark energy density can reach a 1$\sigma$ bias when the equation of state is allowed to vary (top two rows).  The remaining parameters are primarily constrained by Planck.  Their biases are not as significant, although $n_s$ and $A_s$ are biased by almost 0.5$\sigma$ for a flat, \LCDM Universe.  The FoB is not very sensitive to a curvature prior.  The fact that FoB=1.1 indicates that ignoring reduced shear will displace the $w_0-w_a$ error contours by a significant ($\sim 1\sigma$) amount: enough to create an artificial tension with other observations.  Furthermore, the biases may become more significant if combined with other data or priors not considered here.  Since Stage-III surveys will operate in the near future, theoretical systematics may be dominated by other issues such as predicting the matter power spectrum.  Nevertheless, accounting for reduced shear would be the safe approach, and doing so is straightforward via \refeq{RScorrect}.

The biases from neglecting reduced shear are worse for a large survey telescope that would measure a higher density of galaxies and cover more sky than a 4-m class telescope.  As shown in Table \ref{tab:LSSTtable}, dark energy parameters from an LST are consistently biased beyond the 1$\sigma$ level and in some cases exceed a 2$\sigma$ bias.  With FoB=2.5, the true $w_0-w_a$ combination would lie beyond the 2$\sigma$ error contour for those parameters.  Assuming a flat \LCDM Universe (last row) leads to large biases in $n_s$ and $\omega_b$, parameters which can be used to tilt the shear spectra to mimic the reduced shear correction (with $\omega_m$ fixed, $\omega_b$ changes the ratio of baryonic matter to dark matter, which alters early structure growth in a scale-dependent way).  In addition, by the time such a Stage-IV survey is completed, cosmic shear predictions might be accurate enough to make neglecting reduced shear a dominant theoretical systematic error.

\begin{table}
\caption[Estimated biases for a Stage-III survey]{Estimated biases for weak lensing tomography with a Stage-III survey on a 4-m class telescope.  For each parameter, I obtain its expected bias $\Delta p_\alpha$ and its 1-$\sigma$ error $\sigma(p_\alpha)$ from the Fisher matrix, assuming the survey measures shear multipoles $l\le3000$.  Also shown is a dark energy Figure of Bias (FoB) as defined in the text.  In addition to Planck priors (P), I include the following information where indicated:  
fixing $\Omega_k=0$, fixing $w(z)=w_0$, and assuming that dark energy is a cosmological constant ($\Lambda$).}
 \begin{center}
  \begin{tabular}{@{} |c|c|crrrrrrrr| @{}}
    \hline
    Priors & FoB &  &$w_0$&$w_a$&$\Omega_{DE}$&$\Omega_k$&$\omega_m$&$\omega_b$&$n_s$&$\ln A_s$  \\
    \hline
    P & 1.1  & $\Delta p_\alpha\ten{2}$	& -21& 46& 2.2& -0.068& 0.0059 & 0.00065 &0.053& 0.14 \\
     	&  & $\sigma(p_\alpha)\ten{2}$ 	& 24& 64& 2.1& 0.30& 0.12 & 0.017 & 0.61& 1.7  \\
    \hline
    P, $\Omega_k $ & 1.1 & $\Delta p_\alpha\ten{2}$	& -21& 48& 2.2& --& 0.029 & 0.0015 & -0.023& 0.30 \\
     	&  & $\sigma(p_\alpha)\ten{2}$ 		& 24& 64& 2.1& --& 0.060 & 0.014 & 0.52& 1.5  \\
    \hline
    P, $\Omega_k, w$ & -- & $\Delta p_\alpha\ten{2}$	& -3.6& --& 0.92& --& 0.026 & 0.00021 & 0.079& 0.25 \\
     	&  & $\sigma(p_\alpha)\ten{2}$ 		& 4.6& --& 1.2& --& 0.060 & 0.014 & 0.50& 1.5  \\
    \hline
    P, $\Omega_k, \Lambda$ & -- & $\Delta p_\alpha\ten{2}$	& --& --& 0.025& --& 0.015& 0.0031 & 0.21& 0.64 \\
     	&  & $\sigma(p_\alpha)\ten{2}$ 			& --& --& 0.31& --& 0.059 & 0.014 & 0.47& 1.4  \\
    \hline
  \end{tabular}
\end{center}
\label{tab:DEStable}
\end{table}

\begin{table}
\caption[Estimated biases for a Large Survey Telescope]{Same as \reftab{DEStable} but for a Stage-IV Large Survey Telescope.}
 \begin{center}
  \begin{tabular}{@{} |c|c|crrrrrrrr| @{}}
    \hline
    Priors & FoB &  &$w_0$&$w_a$&$\Omega_{DE}$&$\Omega_k$&$\omega_m$&$\omega_b$&$n_s$&$\ln A_s$  \\
    \hline
    P & 2.5  & $\Delta p_\alpha\ten{2}$	& -20& 43& 2.2& -0.078& -0.0047 & 0.0053 &0.31& 0.26 \\
     	&  & $\sigma(p_\alpha)\ten{2}$ 	& 8.8& 24& 0.86& 0.22& 0.10 & 0.016 & 0.56& 1.4  \\
    \hline
    P, $\Omega_k $ & 2.5 & $\Delta p_\alpha\ten{2}$	& -21& 47& 2.2& --& 0.025 & 0.0018 & 0.17& 0.32 \\
     	&  & $\sigma(p_\alpha)\ten{2}$ 		& 8.4& 21& .85& --& 0.057 & 0.013 & 0.41& 1.4  \\
    \hline
    P, $\Omega_k, w$ & -- & $\Delta p_\alpha\ten{2}$	& -2.5& --& 0.67& --& 0.038 & 0.0085 & 0.65& -0.30 \\
     	&  & $\sigma(p_\alpha)\ten{2}$ 		& 2.3& --& 0.51& --& 0.057 & 0.013 & 0.36& 1.4  \\
    \hline
    P, $\Omega_k, \Lambda$ & -- & $\Delta p_\alpha\ten{2}$	& --& --& 0.16& --& 0.0045& 0.015 & 0.90& 0.56 \\
     	&  & $\sigma(p_\alpha)\ten{2}$ 			& --& --& 0.22& --& 0.047 & 0.012 & 0.27& 1.1  \\
    \hline
  \end{tabular}
\end{center}
\label{tab:LSSTtable}
\end{table}
\begin{table}
\caption[Estimated biases for the Joint Dark Energy Mission]{Same as \reftab{DEStable} but for a space telescope survey like the Joint Dark Energy Mission.  The lower rows add cosmic shear information from all modes with $l\le 10^4$.}
 \begin{center}
  \begin{tabular}{@{} |c|c|crrrrrrrr| @{}}
    \hline
    Priors & FoB &  &$w_0$&$w_a$&$\Omega_{DE}$&$\Omega_k$&$\omega_m$&$\omega_b$&$n_s$&$\ln A_s$  \\
    \hline
    P & 1.4  & $\Delta p_\alpha\ten{2}$	& -9.4& 15& 1.5& -0.013& 0.011 & 0.0047 &0.31& 0.52 \\
     	&  & $\sigma(p_\alpha)\ten{2}$ 	& 8.7& 21& 1.0& 0.24& 0.11 & 0.017 & 0.57& 1.5  \\
    \hline
    P, $\Omega_k $ & 1.6 & $\Delta p_\alpha\ten{2}$	& -9.5& 15& 1.5& --& 0.016 & 0.0042 & 0.29& 0.54 \\
     	&  & $\sigma(p_\alpha)\ten{2}$ 		& 8.5& 19& 1.0& --& 0.054 & 0.013 & 0.41& 1.4  \\
    \hline
    P, $\Omega_k, w$ & -- & $\Delta p_\alpha\ten{2}$	& -2.9& --& 0.77& --& 0.025 & 0.0050 & 0.37& 0.29 \\
     	&  & $\sigma(p_\alpha)\ten{2}$ 		& 2.2& --& 0.48& --& 0.053 & 0.013 & 0.40& 1.4  \\
    \hline
    P, $\Omega_k, \Lambda$ & -- & $\Delta p_\alpha\ten{2}$	& --& --& 0.20& --& -0.013& 0.011 & 0.61& 1.6 \\
     	&  & $\sigma(p_\alpha)\ten{2}$ 			& --& --& 0.21& --& 0.045 & 0.012 & 0.36& .97  \\
    \hline
    \hline
    P & 5.2  & $\Delta p_\alpha\ten{2}$	& -20& 31& 3.2& -0.30& -0.12 & 0.026 &1.3& 0.072 \\
     $l\le10^4$	&  & $\sigma(p_\alpha)\ten{2}$ 	& 6.9& 19& 0.72& 0.20& 0.099 & 0.015 & 0.50& 1.4  \\
    \hline
    P, $\Omega_k $ & 5.6 & $\Delta p_\alpha\ten{2}$	& -27& 52& 3.8& --& 0.010 & 0.012 & 0.75& 0.31 \\
     $l\le10^4$	&  & $\sigma(p_\alpha)\ten{2}$ 		& 5.1& 12& 0.62& --& 0.053 & 0.012 & 0.36& 1.4  \\
    \hline
    P, $\Omega_k, w$ & -- & $\Delta p_\alpha\ten{2}$	& -6.7& --& 1.6& --& 0.11 & 0.017 & 1.5& -2.9 \\
     $l\le10^4$	&  & $\sigma(p_\alpha)\ten{2}$ 		& 1.9& --& 0.38& --& 0.047 & 0.012 & 0.32& 1.2  \\
    \hline
    P, $\Omega_k, \Lambda$ & -- & $\Delta p_\alpha\ten{2}$	& --& --& 0.41& --& -0.017& 0.039 & 2.4& 0.32 \\
     $l\le10^4$	&  & $\sigma(p_\alpha)\ten{2}$ 			& --& --& 0.16& --& 0.039 & 0.011 & 0.18& .77  \\   
    \hline
  \end{tabular}
\end{center}
\label{tab:SNAPtable}
\end{table}

Reduced shear will perhaps be an even more important consideration for the Joint Dark Energy Mission, which is expected to have the lowest shape noise of any survey while probing the highest redshifts \citep{JDEM_web}.  The top part of \reftab{SNAPtable} shows that when using modes $l\le 3000$, the dark energy parameter biases can exceed 1$\sigma$.  Biases in $\omega_b$, $n_s$, and $A_s$ are also significant, with the spectral parameters breaking 1$\sigma$ when dark energy parameters are fixed.  Furthermore, by the time JDEM is completed, cosmic shear predictions may be mature enough to exploit even the extremely small-scale modes which can be measured by a space telescope.  The bottom part of \reftab{SNAPtable} assumes that all modes with $l\le 10^4$ will be used in the cosmic shear analysis.  Even with only Planck priors, this assumption leads to very significant biases, including an FoB of 5.2, which corresponds to the correct dark energy parameters mistakenly falling beyond the 99.99\% confidence region of the survey.  It makes sense that including high $l$ modes magnifies the bias because the reduced shear correction increases towards large $l$ and because the errors in the $C_{l;ij}$ decrease as $(2l+1)^{-1/2}$ for uncorrelated, Gaussian modes.  We must keep in mind, though, that the shear modes on non-linear scales are increasingly non-Gaussian and correlated, and therefore the parameter errors calculated here are lower limits (FoBs are upper limits).

\section{Implications for Dark Energy Constraints Using Offset-Linear Scaling}

\subsection{Background and Methodology}

As previously mentioned, cosmic shear is sensitive to large-scale structure formation and to the Universe's expansion history and curvature, a.k.a. the Universe's ``geometry.''  It will be extremely useful to be able to isolate these effects, that is, to use cosmic shear to probe only structure formation or only geometry.  Doing so will allow for consistency checks on cosmic shear data.  Isolating geometry from structure formation will also be useful for testing General Relativity since matter will cluster differently in different gravitational theories, even when the expansion histories match.  Furthermore, predicting the non-linear clustering of matter is a difficult ongoing problem, which is why we must currently impose small-scale cutoffs in cosmic shear analyses; it would therefore be practical and conservative to extract the geometric information from a shear survey in a way that is not biased by our ignorance of the details of non-linear gravitational clustering.  Shear ratio methods do just that - they use cosmic shear to measure the distance/redshift relation, and therefore constrain dark energy, without our having to predict the non-linear matter power spectrum \citep{jain_taylor_2003, zhang_hui_etal_2005, taylor_kitching_etal_2007}.  Although shear ratio methods are useful for the reasons listed above, weak lensing tomography will generally yield tighter parameter constraints since it aggressively incorporates more information.

The shear ratio method I will focus on in this paper is the ``offset-linear scaling'' method of \citet{zhang_hui_etal_2005}.  This method exploits the fact that a cosmic shear cross-spectrum can be written in the following form:
\begin{equation} \label{eq:ratioform}
C_{l;ij} \approx A_{l;i} + \frac{B_{l;i}}{\chieff_j} \:\: \mbox{ for $z_i<z_j$}
\end{equation}
with
\begin{eqnarray}
\frac{1}{\chieff_j} &\equiv& \int d\chi \frac{p_j(z(\chi))}{\ngal_j}\frac{1}{\chi} \label{eq:chieff} \\
A_{l;i} &\equiv& W_0^2\int d\chi' \frac{p_i(z(\chi'))}{\ngal_i} \int_0^{\chi'} \frac{d\chi}{a(\chi)^2}\frac{\chi'-\chi}{\chi'} P_\delta(k,\chi) \\
B_{l;i} &\equiv& -W_0^2\int d\chi' \frac{p_i(z(\chi'))}{\ngal_i} \int_0^{\chi'} \frac{d\chi}{a(\chi)^2}\frac{\chi'-\chi}{\chi'} \chi\, P_\delta(k;\chi)
\:. \end{eqnarray}
Here, I am assuming a spatially flat Universe for simplicity.  Equation \refeq{ratioform} follows directly from \refeq{shearCl} and is not an approximation so long as galaxies are binned by their spectroscopic redshifts.  When photometric redshifts are used, \refeq{ratioform} is valid so long as the galaxies in bins $i$ and $j$ have very little redshift overlap.  The key thing to notice is that by \refeq{chieff}, the effective distance $\chieff_j$ depends only on the distance to the $j$th redshift bin with some weighting given by the galaxy distribution (which is measured).  All of the information about the matter power spectrum $P_\delta$ is contained in the parameters $A_{l;i}$ and $B_{l;i}$.  So it is easy to see that if we choose some common foreground redshift bin, $i=f$, then the following combination of cosmic shear spectra,
\begin{equation} \label{eq:ratiorat}
\Gamma_{abcd}(l;f)\equiv \frac{C_{l;fa}-C_{l;fb}}{C_{l;fc}-C_{l;fd}} = \frac{(\chieff_a)^{-1}-(\chieff_b)^{-1}}{(\chieff_c)^{-1}-(\chieff_d)^{-1}}
\:, \end{equation}
has no dependence on $P_\delta$.  It also has no dependence on $l$ or $f$, but depends only on distances to the background redshift bins.  In other words, we have a cosmic shear observable that is only sensitive to geometric information.  To constrain dark energy, we only have to predict the $\chieff_j$, and we can (in principle) exploit even the non-linear high-$l$ information in a cosmic shear survey since $\chieff_j$ has no dependence on scale.  It is possible to construct observables similar to $\Gamma_{abcd}(l;f)$ using galaxy-shear cross-correlations; however, I will restrict my discussion to cosmic shear.

The form of \refeq{ratioform} implies that any two cross-spectra with a common foreground redshift bin are different only by the scaling of $(\chieff_j)^{-1}$, which is an offset-linear scaling.  As Zhang \etal point out, the scaling relation is invalid if the redshifts from separate bins significantly overlap -- this leads to extra terms in \refeq{ratioform} that do not fit the scaling relation.  Subsequently, the ratio $\shearrat{abcd}{f}{l}$ picks up a dependence on $P_\delta$, thus ruining it as a purely geometric observable.  The extra terms can be made negligible by reducing the scatter in photo-$z$ measurements (a technical challenge) or by simply choosing well-separated redshift bins.  Similarly, since we can only measure reduced shear and not shear, \refeq{ratioform} should contain a term for the reduced shear correction, and again this disrupts the scaling relation.  Since the reduced shear correction always contains a squared factor of the background lensing kernel, $W_j(\chi)^2$, it necessarily produces a term proportional to $(\chieff_j)^{-2}$:
\begin{equation} \label{eq:ratioform2}
C_{l;ij}+\delta C_{l;ij} \approx A'_{l;i}+\frac{B'_{l;i}}{\chieff_j}+\frac{D'_{l;i}}{(\chieff_j)^2} \:\: \mbox{ for $z_i<z_j$}
\:. \end{equation}
Hence a reduced shear cross-spectrum cannot be expressed as an offset-linear scaling relation.

\begin{figure}[tbp]
\begin{center}
\includegraphics[width=5.75in]{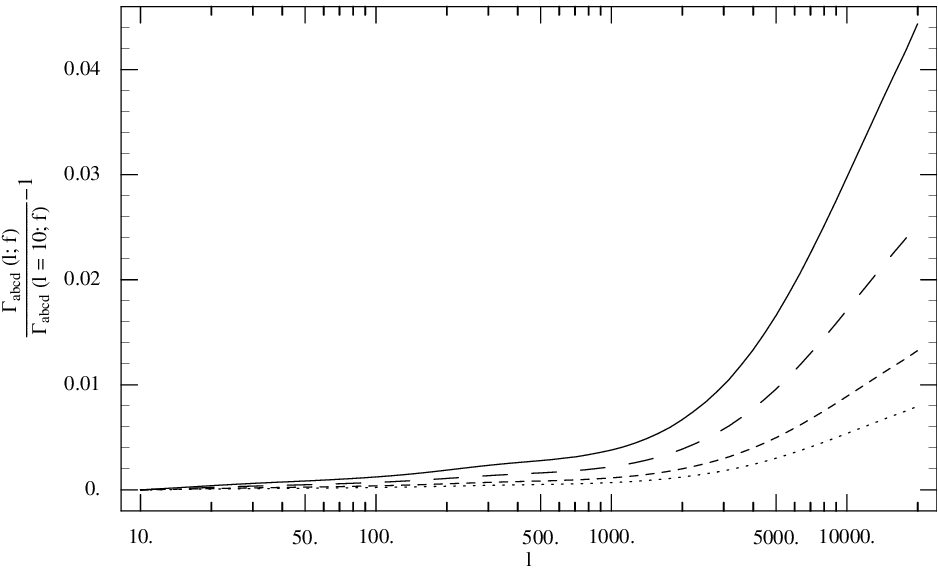}
\end{center}
\caption[Bias in the Shear Ratio Method]{Fractional change in $\shearrat{abcd}{f}{l}$ versus multipole, calculated from \refeq{ratiorat} but including the reduced shear correction.  Curves assume JDEM parameters, and sources have been binned by photometric redshift with bin widths of $\Delta z=0.25$.  The maximum redshifts of the background bins $(a,b,c,d)$ are (3.5, 3.25, 1.25, 1) for the solid curve, (3.5, 3.25, 3.5, 1) for the long-dashed curve, (1.5, 1.25, 1.25, 1) for the short-dashed curve, and (3.5, 1.25, 3.5, 1) for the dotted curve.  All curves have foreground bin $f$ with $0.5\le z<0.75$.  The dependence on $l$ means that the reduced shear cross-spectra do not maintain the offset-linear scaling relation.
\label{fig:RTconstant}  }
\end{figure}

We can see the breakdown of offset-linear scaling in \reffig{RTconstant}, which illustrates the fractional change in $\shearrat{abcd}{f}{l}$ as a function of multipole for a few sample redshift bin combinations.  Without the reduced shear correction, these curves would all be equal to zero, indicating no dependence on $l$ (I have explicitly verified this as a consistency check).  With the correction, the scale independence is clearly broken, and the solid curve shows a combination of redshift bins with one of the largest deviations.  Thus, when the difference between shear and reduced shear spectra are substantial, $\shearrat{abcd}{f}{l}$ is somewhat sensitive to the matter power spectrum and does not provide a purely geometric probe of dark energy.  As previously mentioned, offset-linear scaling also breaks down when photometric redshift bins have substantial overlap, but in that case, we can remove scale dependence e.g. by not using adjacent bins.  It is less clear how to remove the scale dependence introduced by reduced shear, and I propose one possibility in the next section.

To estimate how dark energy parameters are biased by ignoring reduced shear in the offset-linear scaling method, I repeat the Fisher matrix analysis of section \ref{tomography}.  The formalism is essentially the same, but now to make use of the offset-linear scaling of \refeq{ratioform}, I include the $A_{l;i}$ and $B_{l;i}$ as nuisance parameters and marginalize over them so that the observables $C_{l;ij}$ will only be sensitive to $\chieff_j$.
  Clearly, there are now a large number of parameters to keep track of and equally large Fisher and covariance matrices that must be inverted; however, the matrices are sparse, and Zhang \etal describe how to manage them compactly\footnote{For N bins, the Fisher matrix singular if we include the nuisance parameters in the highest two redshift bins ($A_{l;N}$, $B_{l;N}$, $A_{l;N-1}$, and $B_{l;N-1}$) since no combination of $C_{l;ij}$ can measure these.  I remove the singularity by holding these parameters fixed; then $C_{l;N N-1}$ must be ignored to avoid falsely adding information at these redshifts. }  
.  The parameters $\omega_m$, $\omega_b$, $n_s$, and $A_s$ can now be dropped from the Fisher matrix analysis since they do not affect the $\chieff_j$, although I select the same fiducial values in \reftab{cosmopars} to compute the covariance matrices $\Clmat$.  The shear calibration nuisance parameters $f^{\rm cal}_{i}$ can also be discarded since they are completely degenerate with the new nuisance parameters, $A_{l;i}$ and $B_{l;i}$.  I marginalize the Planck Fisher matrix over all but the 3 dark energy parameters (fixing curvature), and I add the resulting matrix as a dark energy prior.  This prior is degenerate, providing no constraints on its own; however it does improve constraints when combined with cosmic shear.

Using the offset-linear scaling method, I repeat the Fisher matrix analysis for the Joint Dark Energy Mission.  All of the cosmological parameters and survey characteristics remain the same as in section \ref{tomography}, except that instead of 5 redshift bins, I now use 16 evenly spaced bins from $0\le z\le 4$.  The purpose of the extra bins is to at least partially make up for the weaker constraints expected from the shear ratio method.  As mentioned previously, finer binning is not proportionally helpful because the shape noise in each bin will be proportionally larger.  For a higher photo-z scatter, wider redshift bins will be needed to avoid substantial spectroscopic overlap as mentioned above.  Since no high-$l$ cutoff is needed, I include all modes with $l<20,000$.

\subsection{Results and Discussion}

As expected, comparing \reftab{ratiotable} with \reftab{SNAPtable} shows that even with finer redshift binning and more small-scale information, dark energy parameter constraints from offset-linear scaling are weaker than those from weak lensing tomography in all cases.  In fact, we cannot put interesting constraints on more than 2 parameters at a time with the offset-linear scaling method and cosmic shear alone.  Although the parameter biases from ignoring reduced shear can be larger than the biases in \reftab{SNAPtable}, they are not as significant.  With $w_a$ and curvature fixed, $w_0$ and $\Omega_{DE}$ are biased by $0.26\sigma$ and $0.27\sigma$ respectively when using modes $l\le20,000$; these are not completely insignificant biases but neither are they a major worry.  Throwing away some high-$l$ modes, where reduced shear is most important, does help to reduce the significance of the bias.  Without modes $l>3000$, constraints on one or two dark energy parameters do not degrade much, yet the biases are considerably smaller.  Hence, decreasing the shape noise should increase bias by providing more weight to information on small scales.  Increasing the sky coverage simply scales up the measurable number of modes in a survey, providing smaller error bars on dark energy parameters; since $\Delta p_\alpha$ is nearly independent of $\fsky$ in the shear ratio method, $\Delta p_\alpha/\sigma(\alpha_p)$ scales roughly as $\sqrt{\fsky}$ (not a perfect proportion when including the Planck prior).  Thus, I find that if we expand the JDEM survey to $\fsky =0.5$ and fix $w_a$, the $0.26\sigma$ bias on $w_0$ becomes $0.59\sigma$.
\begin{table}[tbp]
\caption[Estimated biases for the shear ratio method with JDEM]{Same as \reftab{DEStable} but using the shear ratio method with a space telescope survey like the Joint Dark Energy Mission.}
 \begin{center}
  \begin{tabular}{@{} |c|c|crrr| @{}}
    \hline
    Priors & FoB & ¥ & $w_0$ & $w_a$ 	& $\Omega_{DE}$ \\
    \hline
    P, $\Omega_k $ & 0.35	& $\Delta p_\alpha\ten{2}$	& -47	& 120	& 4.2 \\
    $l\le3000$ 	& ¥ 				& $\sigma(p_\alpha)\ten{2}$ 		& 140	& 360	& 12 \\
    \hline
    P, $\Omega_k, w$ & -- 	& $\Delta p_\alpha\ten{2}$	& -1.8	& -- 	& 0.49\\
    $l\le3000$ 	& ¥ 				& $\sigma(p_\alpha)\ten{2}$ 		& 16	& -- 	& 4.6	\\
    \hline
    P, $\Omega_k, \Lambda$ 	& -- & $\Delta p_\alpha\ten{2}$	& --		& --		& 0.0014	\\
    $l\le3000$ 	& ¥ 				& $\sigma(p_\alpha)\ten{2}$ 			& --		& --		& 0.66	\\
    \hline
    \hline
    P, $\Omega_k $ & 0.52	& $\Delta p_\alpha\ten{2}$	& -53	& 150	& 3.5 \\
    $l\le20,000$ 	& ¥ 				& $\sigma(p_\alpha)\ten{2}$ 		& 130	& 340	& 11 \\
    \hline
    P, $\Omega_k, w$ & -- 	& $\Delta p_\alpha\ten{2}$	& 3.9	& -- 	& -1.1 \\
    $l\le20,000$ 	& ¥ 				& $\sigma(p_\alpha)\ten{2}$ 		& 15	& -- 	& 4.1	\\
    \hline
    P, $\Omega_k, \Lambda$ 	& -- & $\Delta p_\alpha\ten{2}$	& --		& --		& -0.019	\\
    $l\le20,000$ 	& ¥ 				& $\sigma(p_\alpha)\ten{2}$ 			& --		& --		& 0.66	\\
    \hline
  \end{tabular}
\end{center}
\label{tab:ratiotable}
\end{table}

A potential advantage of the offset-linear scaling method is its ability to use high-$l$ modes without our having to predict shear spectra for those modes.  The importance of reduced shear means that in order to exploit these modes without bias, the method will have to be adjusted.  When using weak lensing tomography, reduced shear can be accounted for by simply including the reduced shear correction $\delta C_{l;ij}$, given by \refeq{RScorrect}, in our predictions.  Unfortunately, since shear ratio methods do not require predictions of shear spectra, only the distances $\chieff_j$, there is no prediction that can be corrected.  To be clear, I have calculated cosmic shear spectra in this section for the purposes of forecasting, but to do this, I had to explicitly assume an underlying matter power spectrum.  In practice, a shear ratio analysis will not depend on such information.  One could conceivably guess the correction using a reasonable bispectrum prediction; however, for consistency, we should de-bias the shear ratio method in a way that does not involve calculating the reduced shear correction from theory.  One option is to add an extra set of nuisance parameters, such as $D'_{l;i}$ in \refeq{ratioform}, to account for the deviation from an offset-linear scaling relation.  However, this will likely degrade dark energy constraints to the  point of being uninteresting.  Another option is to {\em measure} a reduced shear correction rather than calculate it.  Equation \refeq{RScorrect} tells us that the needed correction is an integral over all modes of the convergence bispectrum, $B^\kappa$.  Hence, one could conceivably use measurements of shear bispectra to infer the correction and then convert reduced shear spectra to shear spectra, thereby recovering the offset-linear scaling form.   Bispectra should be readily available from weak lensing surveys since they are useful in their own right for studying non-Gaussianity and non-linear clustering \citep{cooray_hu_2001a, takada_jain_2004}.

\section{Conclusions and Caveats}
\label{cha:conclude}

Reduced shear, the lensing observable measured by averaging galaxy ellipticities, is typically approximated by shear for calculational convenience.  In this paper, I have explored the error due to this approximation in the context of cosmic shear power- and cross- spectra. I have presented the general formula, \refeq{RScorrect}, for converting shear spectra to reduced shear spectra, and the 2D version of this perturbative correction was found to agree with simulations in previous work \citep{dodelson_shapiro_etal_2006}.  The correctional term is a projection of the matter bispectrum, which is enhanced by non-linear gravitational clustering; the correction is therefore most significant on small scales and for highly clustered matter fields, for instance when $\sigma_8$ is large.

I have investigated the extent to which ignoring reduced shear will bias 3D weak lensing methods - methods that measure sheared galaxies at multiple redshifts in order to extract information about the expansion of the Universe and the growth of large-scale structure.  Weak lensing surveys that constrain dark energy parameters via tomography can neglect reduced shear in the short term because it is unlikely to be the dominant theoretical systematic error.
However, as the finer details of shear maps come into focus -- observationally and theoretically -- reduced shear will almost certainly need to be accounted for, as I have demonstrated using a Fisher matrix analysis.  I have also shown that the shear ratio method known as offset-linear scaling, which probes dark energy by isolating the geometric information in cosmic shear, becomes invalid on small angular scales because of reduced shear.  This method will probably need to be adjusted in order to take advantage of information on the smallest angular scales of shear maps.

The tables in this paper that compare biases in cosmological parameters to their forecasted uncertainties are only intended to be a rough guide -- they should not be considered definitive indicators of the importance of reduced shear, particularly since a real survey will not have the perfectly Gaussian likelihood functions assumed by a Fisher matrix analysis.  The actual characteristics of the shear surveys I considered will also determine the importance of the discrepancy; in particular, photometric redshift uncertainties will degrade parameter constraints and could change the significance of bias.  Moreover, combining weak lensing results with different priors or with other dark energy probes (such as type Ia supernovae) could make a parameter bias better or worse, depending on the direction of the bias in parameter space and any tension between the likelihood functions being combined.  Lastly, the actual difference between shear and reduced shear spectra depends strongly on the underlying matter power spectrum, as illustrated by the sensitivity of the correction to $\sigma_8$.

As a final note, cosmic shear is not the only lensing observable to which reduced shear is applicable.  Most notably, shear ratio methods are powerful when applied to galaxy-shear cross correlations.  On small angular scales, we should expect galaxy-shear cross spectra to deviate from the offset-linear scaling form in a way similar to the shear-shear deviations plotted in \reffig{RTconstant}.  Furthermore, shear bispectra and shear-galaxy bispectra may eventually require a 4th order reduced shear correction \citep{dodelson_zhang_2005}.

\acknowledgments{

I am grateful to Michael Turner and Scott Dodelson for their guidance on matters both scientific and professional and to Michael for encouraging me to see ``the big picture.''  Thanks also to Bruce Winstein and Jeff Harvey for being on my Ph.D. committee and ensuring that I graduate with a quality dissertation.  I was lucky to have Nobuko McNeill looking out for me at the University of Chicago, making my graduate career go as smoothly as possible.  Thanks to Neal Dalal for suggesting that I take a look at reduced shear in shear ratio methods, and to David Bacon, Robert Crittenden, Simon DeDeo, Felipe Marin, Fabian Schmidt, and Martin White for useful discussions.  Thanks also to Andrew Zentner for the use of his matter power spectrum code, and to Jeremy Tinker for his assistance with galaxy bias calculations that ultimately were not included in this work.

This work was supported in part by the Department of Energy and by the Kavli Institute for Cosmological Physics through grants NSF PHY-0114422 and NSF PHY-0551142 and an endowment from the Kavli Foundation.  CS is currently supported by the UK Science and Technology Facilities Council.
}


\newpage
\bibliography{master_library}
\bibliographystyle{hapj}


\end{document}